\documentclass[aps,preprintnumbers,floats, prd,nofootinbib]{revtex4-1}

\usepackage{epsfig}
\usepackage{amssymb}
\usepackage{bm}
\usepackage{hyperref}
\usepackage{color}
\usepackage{amsmath}

\newcommand{\be}{\begin{equation}}
\newcommand{\ee}{\end{equation}}
\newcommand{\bea}{\begin{eqnarray}}
\newcommand{\eea}{\end{eqnarray}}

\renewcommand{\slash}{\displaystyle{\not}}

\newcommand{\nn}{\nonumber}

\newcommand{\met}{$\slash{E}_T$~}

\begin{document}

\begin{flushright}
MIFPA-14-01 
\end{flushright}

\title{\bf\Large Probing Light Nonthermal Dark Matter at the LHC}

\author{Bhaskar Dutta$^1$}
\email{dutta@physics.tamu.edu}
\author{ Yu Gao$^1$}
\email{yugao@physics.tamu.edu}
\author{Teruki Kamon$^{1,2}$}
\email{kamon@physics.tamu.edu}
\affiliation{\bigskip 
$^{1}$~Mitchell Institute for Fundamental Physics and Astronomy, \\
Department of Physics and Astronomy, Texas A\&M University, \\ College Station, TX 77843-4242, USA \\
$^{2}$~Department of Physics, Kyungpook National University, \\ Daegu 702-701, South Korea}

\begin{abstract}
This paper investigates the collider phenomenolgy of a minimal nonthermal dark matter model with a 1-GeV dark matter candidate, which naturally explain baryongensis. Since the light dark matter is not parity-protected, it can be singly produced at the LHC.
This leads to large missing energy associated with an energetic jet whose transverse momentum distribution is featured by a Jacobian-like shape. 
The monojet, dijet, paired dijet and 2 jets + missing energy
channels are studied. 
Currently existing data at Tevatron and LHC offer significant bounds on our model.
\end{abstract}

\maketitle

\section{Introduction}

Weakly interacting massive particles (WIMPs) are very promising dark matter (DM) candidates~\cite{WIMP} which arise in models beyond the standard model (SM). The lightest supersymmetric particle in supersymmetry models is one of the best examples of such a particle. The current LHC bounds on colored sectors, however, have put these models under  pressure. The limits on the mass of gluinos and squarks of the first generation is getting higher which has motivated new scenarios,  e.g., natural SUSY~\cite{Natural} where the Higgsino typically serves as the DM candidate. The annihilation rate for sub-TeV Higgsinos is larger than  the annihilation rate required in the thermal scenario~  $\left<v\sigma\right>=3\times 10^{-26}$ cm$^3$/sec.  Consequently, non-thermal mechanism and/or more than one DM candidates is needed to obtain the correct DM abundance. Large annihilation rates are, however, constrained by the Fermi-LAT data from DM annihilation~\cite{Fermi}. For smaller dark matter mass, this data  prefers smaller (compared to the thermal case) annihilation rate which may result into over-abundance dark matter scenarios and requires non-thermal mechanism for correct relic abundance. 

Further, constraints on the top squark mass have also put the electroweak baryogenesis~\cite{EW} in the minimal supersymmetric standard model (MSSM) in a tight corner. We need alternate scenarios to explain the current explanation of the baryon asymmetry of the universe~\cite{Baryo}. e.g.,  leptogenesis~\cite{Lepto}, Affleck-Dine baryogenesis~\cite{AD}, hidden sector baryogenesis~\cite{Hidden}, etc. 

 Recently  a minimal extension of the SM was presented to address the above mentioned issues~\cite{Allahverdi:2013mza}. New renormalizable baryon number violating interactions was introduced in the Lagrangian that can lead to a successful baryogenesis. The minimal field content that is required to achieve this includes iso-singlet color-triplet scalars ($X$) and one singlet Majorana fermion ($n_{DM}$). Here $n_{DM}$ becomes stable, hence a DM candidate, when its mass is around ${\cal O}({\rm GeV})$.
Both the baryon abundance and the DM content of the universe can therefore be motivated from the simplest extension of the SM Lagrangian.
In such an extension,
the DM relic density and the baryon asymmetry are produced non-thermally from the decay of some heavy particle(s)~\cite{Allahverdi:2012gk,Allahverdi:2010im} even though the dark matter annihilation rate is much smaller than the thermal annihilation rate. One interesting point about this scenario is that since the DM mass is ${\cal O}({\rm GeV})$, a correlation between the number densities  automatically translates into a similar relation between the DM and baryon energy densities  which can provide a natural explanation of the baryon-DM coincidence puzzle~\cite{Coincidence}. 

In this model, the DM candidate interacts with up-type quarks via the exchange of colored scalar fields. This new colored fields also interacts with the down type quarks. It was shown that that the resulting spin-independent and spin-dependent DM-nucleon scattering cross sections are well below the bounds from current and upcoming experiments, which makes the prospects for direct detection weak. The small annihilation cross-section make the prospects for indirect detections weak as well.

However, the model may be probed at the LHC via the colored scalars if they have ${\cal O}({\rm TeV})$ masses. The  DM candidate $n_{DM}$ which is a source of missing energy can be produced singly along with a quark at the LHC in this model which makes it different from the DM scenarios where the dark matter are pair produced~\cite{bib:othermonojet}, e.g. under the protection of certain discrete symmetries. In such models, the energetic jet in the monojet (plus missing energy) signal arises from the initial state radiation, or the decays of pair-produced heavy particles.  In this paper, we identify the possible signals of this minimal extension of the SM at the LHC, which includes, monojet, dijet, paired dijet and 2 jets + missing transverse energy($\slash{E}_T$) channels. The existing analysis on these final states have already started putting constraints on the parameter space of this model.

This paper is organized as follows: Section~\ref{min_model} presents a minimal implementation of the model for collider phenomenology. In Section~\ref{sect:pheno} we discuss the various channels that can be studied at the LHC. Section~\ref{sect:results} show the constraints from current collider data. We conclude and briefly comment on possible experimental search improvements and additional channels in Section~\ref{sect:disc}.

\section{A minimal model}
\label{min_model}

The interaction Lagrangian is,
\be {\cal L}_{int}=
\lambda_1^{\alpha,\rho\delta}\epsilon^{ijk} X^{}_{\alpha,i} \bar{d}^c_{\rho,j} \text{\bf{P}}_R d_{\delta,k} + \lambda_2^{\alpha,\rho} X^*_{\alpha} \bar{n}_{\text{DM}\rho} \text{\bf{P}}_R u +  \text{C.C.} \label{eq:L_int}
\ee
where $d^c$ is the charge-conjugate of the Dirac spinor. $\text{\bf{P}}_R$ is the right-handed projection operator. $X$s are iso-single color triplet scalars with hypercharge 4/3 and $n_{DM}$ is a SM singlet which is dark matter candidate in this model. For the indices, $\rho,\delta=\{1,2,3\}$ denote the three quark generations, and $i,j=\{1,2,3\}$ are the SU(3) color indices. Successful baryogenesis requires more than one new scalar~\cite{Allahverdi:2013mza}, thus $\alpha=1,2$ denotes for a minimal case with two $X$ fields.

In principle the flavor indices in $\lambda$ allow a large number of free parameters, plus their complex phases~\cite{Allahverdi:2013mza} that give CP violation in the early universe. For collider searches, we focus on a simplified case where we ignore the flavor struture and write the coupling coefficients as,
\be 
\lambda_1^{\alpha,\rho\delta}=\lambda_1\cdot \lambda^{\alpha}_{1X} \cdot \lambda^{\rho\delta}_{1R},
\ee
where a single real $\lambda_1$ sets the overall scale of coupling strength. For the $X$ and quark generation structures, we naively assign as,
\be 
\lambda^{\alpha}_{1X}= (1,1)~\text{\hspace{1cm}and\hspace{1cm}}~
\lambda^{\rho\delta}_{1R}=\left(
\begin{array}{ccc}
0&1&1\\  0&0&1\\ 0&0&0\\
\end{array} \right).
\label{eq:struct1}
\ee
Note the $\lambda^{\rho\delta}_{1R}$ can only maintain its antisymmetric component, due to the antisymmetric structure in the $SU(3)$ color indices. Similarly, 
\be 
\lambda_2^{\alpha,\rho}=\lambda_2\cdot \lambda^{\alpha}_{2X} \cdot \lambda_{2R}^{\rho},
\ee
where
\be 
\lambda^{\alpha}_{2X}=(1,1) ~\text{\hspace{1cm}and\hspace{1cm}}~
\lambda^{\alpha}_{2R}=(1,1,1).
\label{eq:struct2}
\ee
Here all three generations share the same coupling. All the complex phases in these parameters are dropped here as they only appear in the interference terms at loop level between two different $X$s. For collider searches, s-channel tree-level diagrams dominate and Eqs.~\ref{eq:struct1} and~\ref{eq:struct2} suffice if two $X$ are not extremely degenerate in mass to cause interference. When interference between $X1$ and $X2$ occurs, the complex phases cannot be neglected; to simplify our collider study, we identify several general scenarios in Appendix~\ref{sect:scenarios}. The interference can be negligible when  $|\lambda_1|\sim |\lambda_2|$ or $|\lambda_1|\gg |\lambda_2|$. However, $|\lambda_1|\ll |\lambda_2|$, $X1-X2$ interference generally occurs and collider bounds may become very sensitive to model parameters.

At this stage, the interaction terms are described by a set parameter set $\{ \lambda_1, \lambda_2, M_{X1}, M_{X2}\}$, i.e. two scalar couplings and the mass of two $X$ fields. As shown in the next section, the lighter one of $X$s dominates the cross-section in most cases. Without losing generality, we assume $X1$ be lighter and set $M_{X2}=2M_{X1}$ in the rest of this paper.  By doing so we obtain the collider bounds on the leading contributor; when $X1$ and $X2$ are closer in mass (but still no interference), the combined bounds on cross-section can be simplify scaled a factor of 2. When interference occurs in the $|\lambda_1|\ll |\lambda_2|$ region, a strong bound by a factor of 4 serves as the most optimistic constraint. We use the {\it FeynRules}~\cite{Christensen:2008py} 
software to implement this minimal model as a {\it Madgraph5}~\cite{Alwall:2011uj} package.

\section{Collider phenomenology}
\label{sect:pheno}

The most striking feature of the model's signal is that the dark matter can be singly produced in the decay of a heavy colored scalar $X$. This results in distributions of jet $p_T$ (and $\slash{E}_T$ in the monojet case) exhibiting peaks at half of the mass of $X$, as discussed later.% 

We investigate possible collider signals that are can be tested at the LHC, and categorize $X$ production mechanisms on the number $X$s in the hard scattering process. Monojet and dijet events are occur via an s-channel resonance, while channels with multiple jets + \met, as well as the paired dijet channel, receive significant contribution from $X$ pair production.

\subsection{Single-$X$ channels}

\begin{figure}[h]
\includegraphics[scale=0.6]{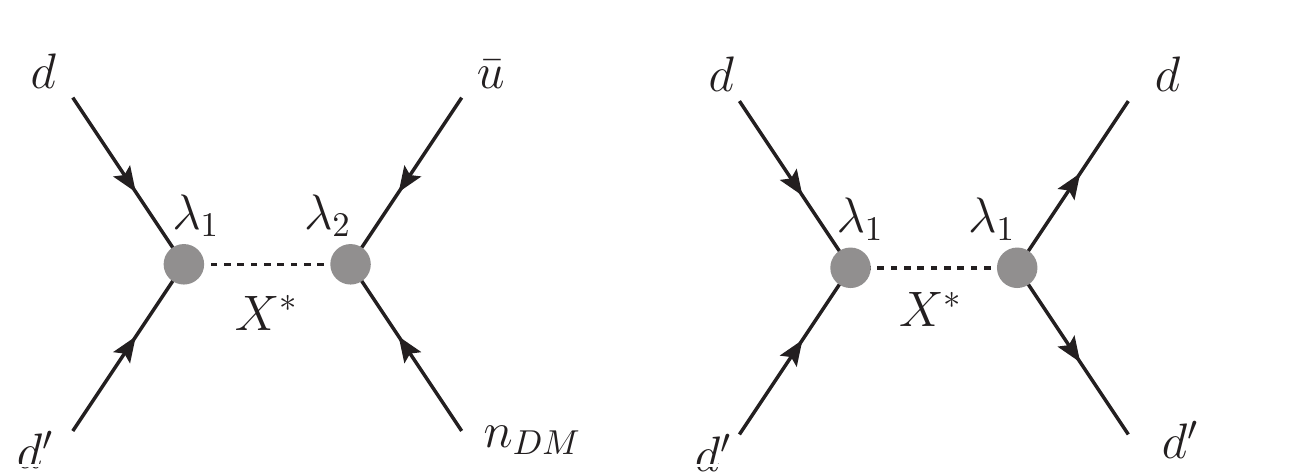}
\caption{Feynman diagrams leading to monojet (left) and dijet (right) final states at the LHC.}
\label{fig:feynmanDiagram}
\end{figure}

The monojet channel occurs via a s-channel $X$ resonance. The monojet's Feynman diagram is shown in Fig.~\ref{fig:feynmanDiagram} (left). The jet recoils against the missing particle $n_{DM}$ and its transverse moment peaks near one half of the resonance energy $\sqrt{\hat{s}}=M_{X1}$. This is illustrated later in Fig.~\ref{fig:pt_distr}. The signal cross-section does not suffer from a high $p_T$ cut in monojet searches. In contrast, in models where dark matter must be pair produced, monojet events arise from initial state radiation (ISR), and the jet $p_T$ would peak at low energy due to infrared and collinear divergences.

The PDF-integrated total cross-sections scale as $\sigma \propto {|\lambda_1|^2|\lambda_2|^2}/{\Gamma_{X1}}$, and the
$X1$ decay width is given by,
\be 
\Gamma_{X}=\frac{1}{8\pi M^2_X}
\left[2|\lambda_1|^2\sum_{i\neq j}|\vec{p}_{ij}|(M_X^2-M^2_{d_i}-M^2_{d_j})
+|\lambda_2|^2\sum_{i}|\vec{p}_{i}|(M_X^2-M^2_{u_i}-M^2_{n_{DM}})\right].
\ee
The $u$ and $d$ denotes for any up/down-type quarks. The $\vec{p}_{ij}$ is the final state momentum which depends on the mass of final state particles, e.g. $M_{d_i}$ and $M_{d_j}$, where indices $\{i,j \}=1,2,3$ denote for different quark generations. Similar $\vec{p}_{i}$ is the final state momentum for the $X\rightarrow u_i~n_{DM}$ decay.
In the heavy $M_X$ limit, $\sigma \propto {|\lambda_1|^2|\lambda_2|^2}/(2|\lambda_1|^2+|\lambda_2|^2)$. This parameter dependence makes the monojet cross-section into two regions:

  ({\it i}) $\lambda_1\approx \lambda_2 \equiv \lambda$, where  $\sigma\propto |\lambda|^2$.
  
  ({\it ii}) $\lambda_1 \ll \lambda_2$ or $\lambda_1 \gg \lambda_2$, where the $X$ width becomes dominated by the larger of $\lambda_1 , \lambda_2$, which cancels itself in the numerator and $\sigma\propto |\text{min}(\lambda_1,\lambda_2)|^2$.
 
 It can be generalized that the monojet cross-section is determined by the lesser of $\lambda_1$ and $\lambda_2$. In the next section, we will show the LHC's constraint in both cases .
 
\begin{figure}[h]
\includegraphics[scale=0.6]{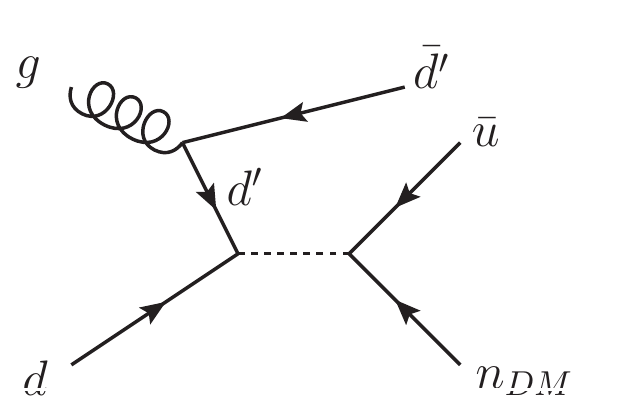}
\caption{The ISGS diagram that leads to a 2 jets +$\slash{E}_T$ final state at the LHC. Here $d$ and $d'$ are of different down-type quark generations if connected by the same $\lambda_1$ vertex.}
\label{fig:ISGS}
\end{figure} 
 
 The dijet diagram is shown in the right panel of Fig.~\ref{fig:feynmanDiagram}. This channel potentially offers a complementary constraint on only $\lambda_1$, or  $\sigma \propto {|\lambda_1|^4}/(2|\lambda_1|^2+|\lambda_2|^2)$. Compared to the monojet case, $\lambda_2$ is almost irrelevant unless it is larger than $\lambda_1$ and dominates the $X$ scalar width. We investigate this channel with the CDF~\cite{Aaltonen:2008dn} dijet data due to its lower dijet mass threshold, and superior constraint compared to currently available LHC results.

When multiple jets and missing energy are both considered, the leading contributor is the initial state gluon-splitting (ISGS) diagrams, as shown in Fig.~\ref{fig:ISGS}, which gives a two jets plus $\slash{E}_{T}$ final state that can be testes with existing LHC searches. In contrast to mono/di-jet cases, this process benefits from the valence $d$-quark and gluon not being PDF suppressed and have a sizeable cross-section at the LHC. Initial state radiations can be added to diagrams in Fig.~\ref{fig:feynmanDiagram}, but their contribution is limited to due high jet $p_T$ cuts in multijet+$\slash{E}_{T}$ search channels.

\subsection{ Two-$X$ channels}

\begin{figure}[h]
\includegraphics[scale=0.6]{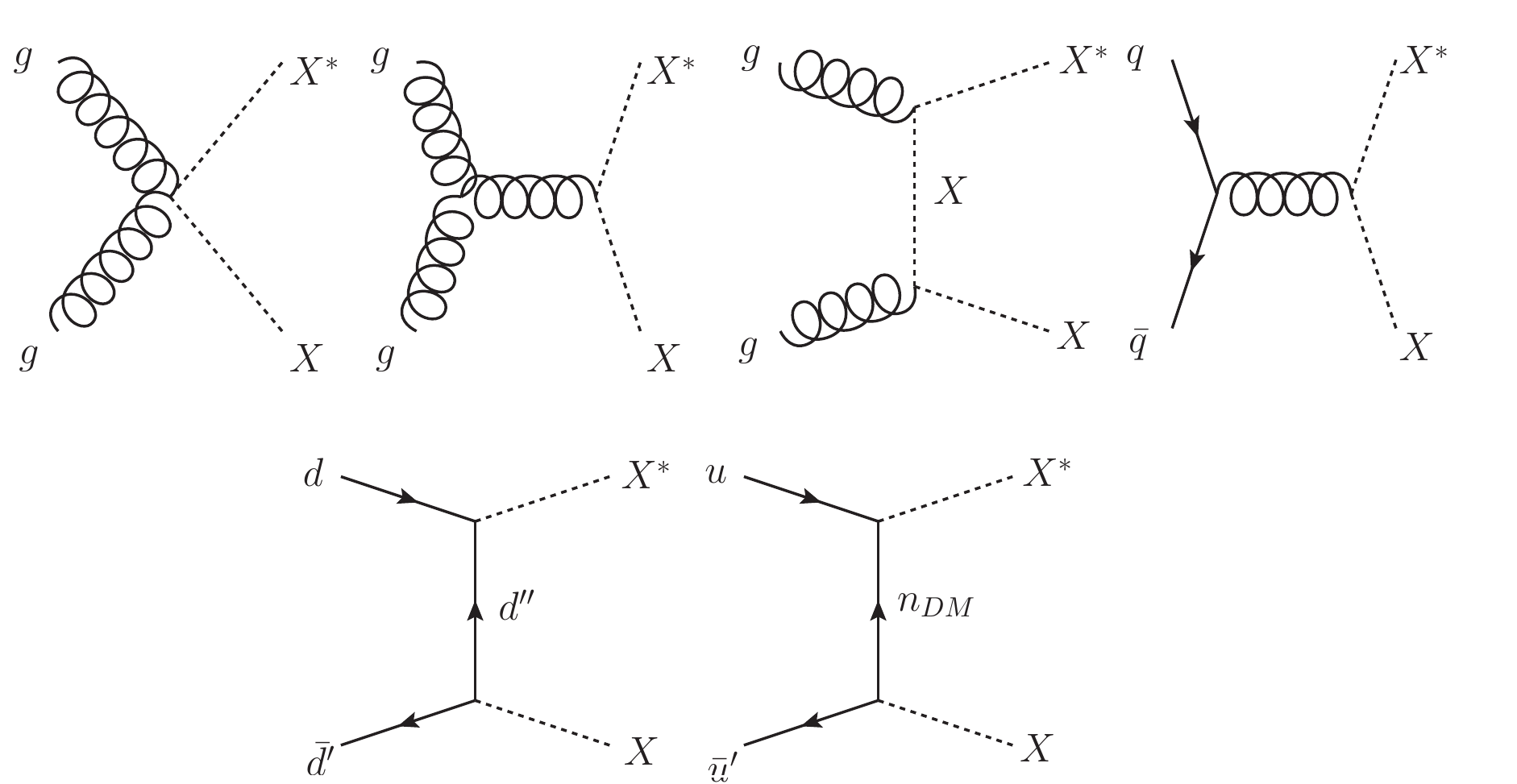}
\caption{Feynman diagrams for the pair production of $X$. Here $d,d'$ and $d''$ must be of different quark generations if connected by the same $\lambda_1$ vertex.}
\label{fig:pair_feynman}
\end{figure}

The pair production can rise from both QCD and the new physics (NP) vertices given by Eq.~\ref{eq:L_int}. The relevant Feynman diagrams are illustrated in Fig.~\ref{fig:pair_feynman}.

Out of the five diagrams, the latter two dominate the pair production at comparable coupling strength because of their very light $t$-channel exchange particles. Their contributions scale with $\lambda_2^4$ or $\lambda_1^4$, respectively. In comparison the QCD contribution is independent from $\lambda_1$, $\lambda_2$, and can become important at low $\lambda$ values, e.g. in case of a tight experimental bound.

As $X$ can decay either into one jet with missing energy ($X\rightarrow u~ n_{DM}$), or  two jets ($X\rightarrow d d'$), our model can also be tested by following channels at the LHC:

(1) two (or three) jets + $\slash{E}_T$, with both (one) $X$ decay into $u,n_{DM}$;

(2) two pairs of  dijets, with both $X$s decay into $d,d'$.

In this study, we calculate signal rates at parton-level, and only consider two jets + $\slash{E}_T$ in case (1), as the lowest order for the multijet+ $\slash{E}_T$ search channel. 

While $\lambda_1$ and $\lambda_2$ play symmetric roles during the pair-production, a larger $\lambda_1$ raises the $X$ decay branching fraction into $dd'$ hence enhances Channel (2), while a larger $\lambda_1$ leans towards Channel (1). Therefore, these two channels, if dominated by pair-production diagrams\footnote{The ISGS contribution to two jets + $\slash{E}_T$ is determined by the lesser between $\lambda_1$ and $\lambda_2$.}, can give complimentary constraints on both $\lambda_1$ and $\lambda_2$.

Since we carry out signal calculations at the parton level, to compare Channel (1) with experimental results, it is necessary to adopt 2 jets + $\slash{E}_T$ exclusive data from ATLAS~\cite{ATLAS-CONF-2013-047}. For Channel (2), we test against the paired-dijet results from CMS~\cite{CMS-PAS-EXO-11-016} and two $X$ masses can be reconstructed.

\section{Collider constraints}
\label{sect:results}

Here we show the collider constraints from each search channel in the previous section. As the $X$ mass is expected to be around the TeV scale to explain the relic density, we use two benchmark points, $M_{X1}=500$ GeV and 1 TeV, to calculate the signal cross-section and compare with new physics bounds in each relevant channel. The constraints are plotted on the parameter space $\{\lambda_1,\lambda_2\}$. All signals are generated at the parton level with the {\it Madgraph5}, and the (anti)proton PDF(s) assume {CTEQ6l}~\cite{CTEQ6}. Note: in this paper, we do not require $b$-tagging and count a parton-level $b$ quark as a jet.

\begin{figure}[h]
\includegraphics[scale=0.6]{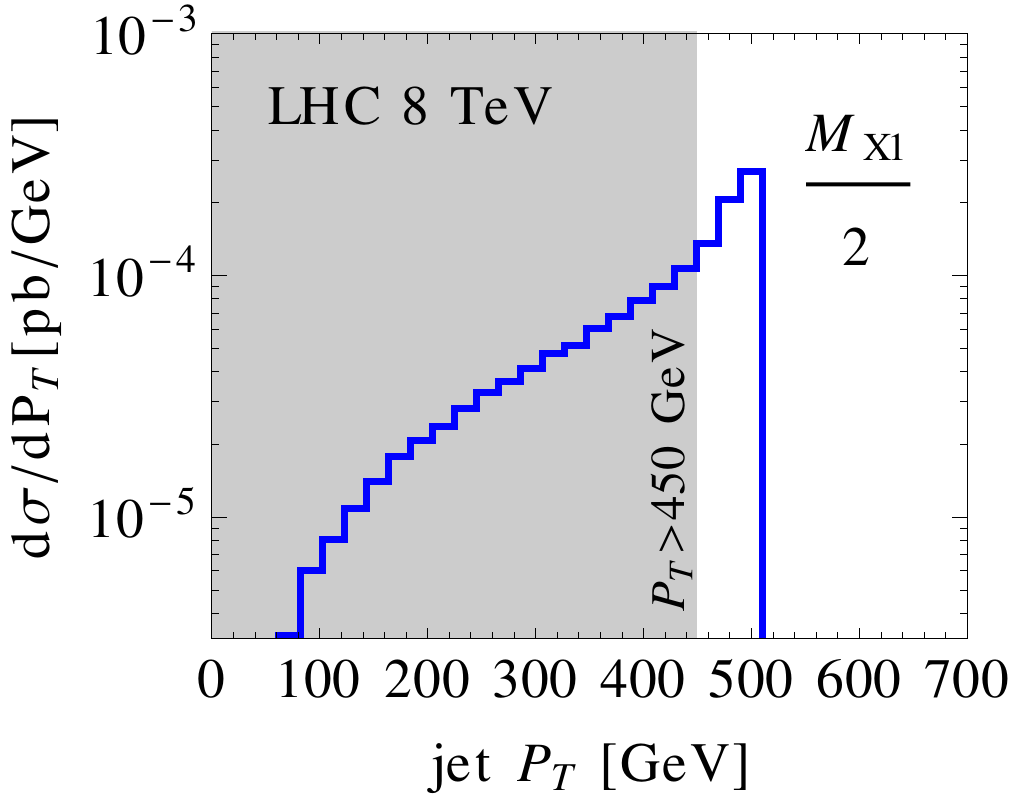}
\caption{Monojet $p_T$ distribution for $M_{X1}$=1 TeV. Among all the $p_T$ cuts in Ref.~\cite{CMS-PAS-EXO-12-048}, the 450 GeV cut is the closest to $M_{X1}/2$ and gives the most stringent constraint.}
\label{fig:pt_distr}
\end{figure}

For monojet+$\slash{E}_T$, the visible jet recoils against the missing momentum, hence the jet $p_T=\slash{E}_T$. 
As illustrated in Fig.~\ref{fig:pt_distr}, the distribution of jet transverse momentum is featured by two Jacobian-like  peaks near one half of the resonance energy $\sqrt{\hat{s}}=M_{X1}$ and $\sqrt{\hat{s}}=M_{X2}$. The transverse mass of the leading jet $p_{T}$ and MET infers the mass of $X1$ and provide a maximal signal significance.

\begin{figure}[h]
\includegraphics[scale=0.6]{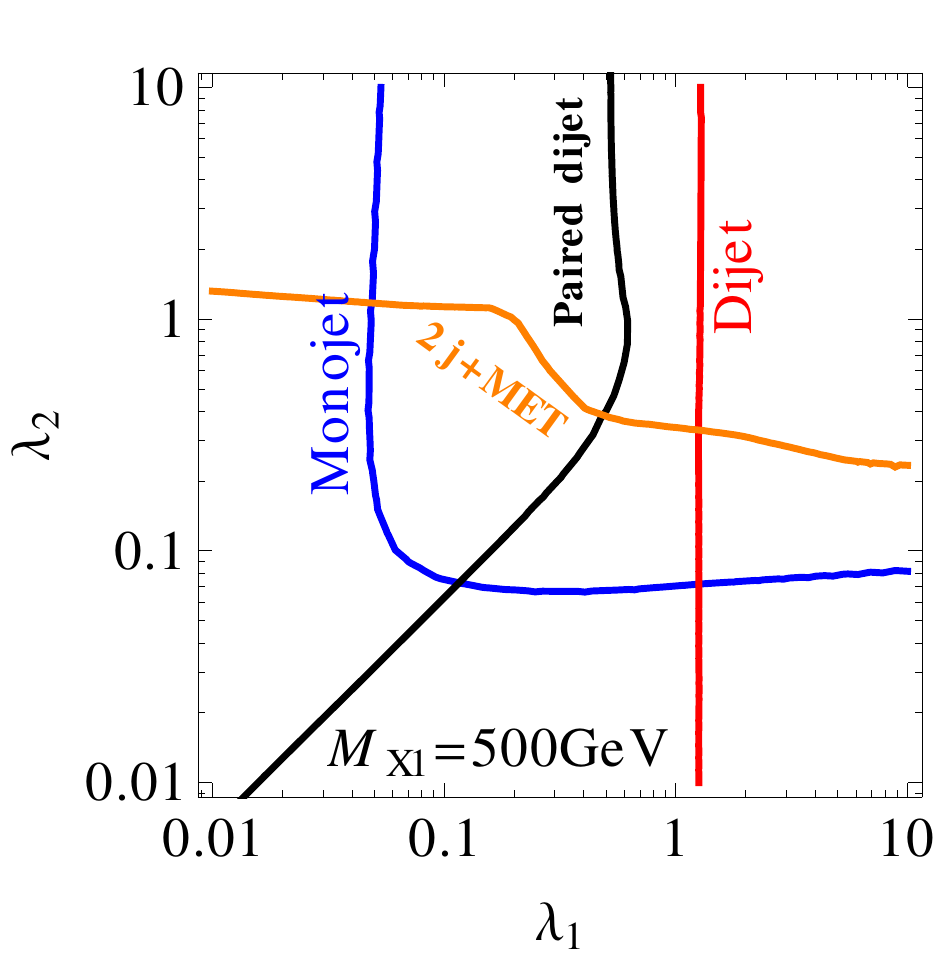} \hspace{0.1cm}
\includegraphics[scale=0.6]{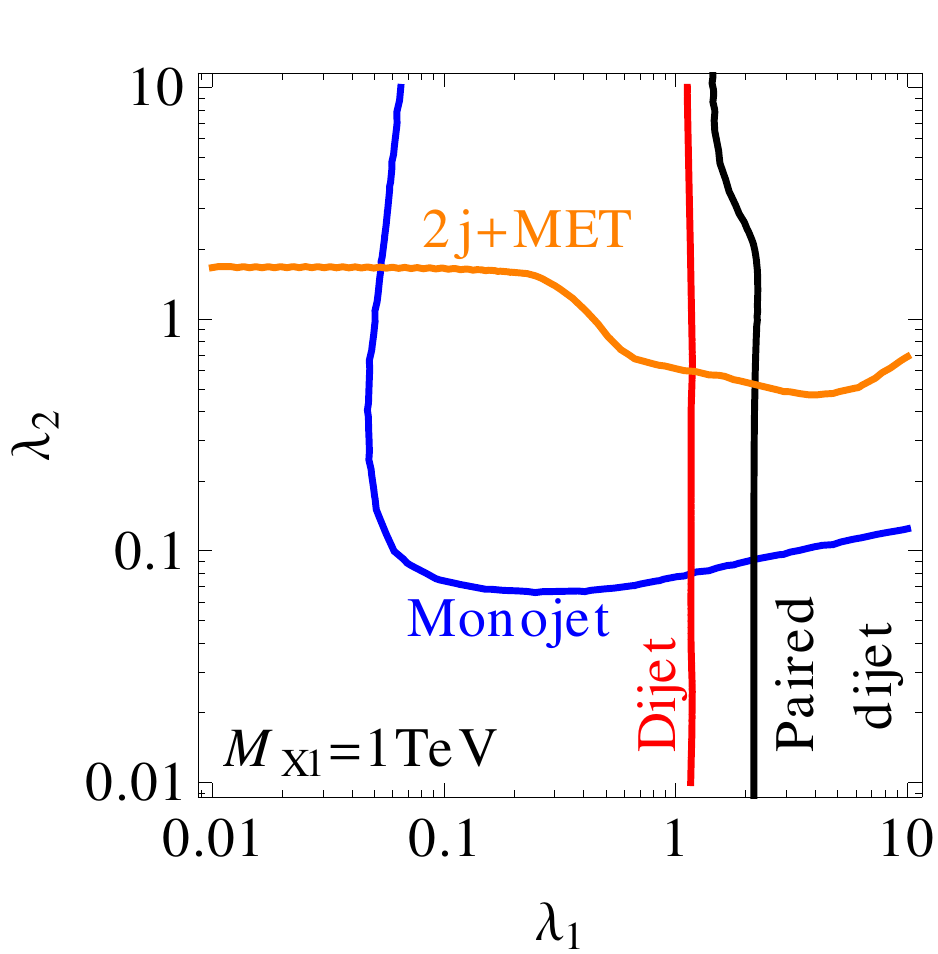}
\caption{A collection of bounds from current collider results. See the discussion in Section~\ref{sect:results} for the data set and kinematic cut used for each channel.}
\label{fig:bounds}
\end{figure}

For monojet events, we require the jet psedurapidity $|\eta_j|<2.4$ and the various threshold jet $p_{T}$ cuts (250 GeV to 550 GeV) listed in the Ref.~\cite{CMS-PAS-EXO-12-048} for 20 fb$^{-1}$ data at 8 TeV. We try all these cuts and select the most stringent one on $\lambda_1$ and $\lambda_2$. The 95\% credence level (C.L.) bounds on $\{\lambda_1,\lambda_2\}$ at the benchmark $X1$ masses are shown as blue curves in Fig.~\ref{fig:bounds}. As discuss in the previous section, the monojet cross-section depends on the lesser of $\lambda_1, \lambda_2$ if one of them is much smaller than the other. At both benchmark points, the smaller $\lambda$ is constrained to $\sim {\cal O}(0.1)$. This bound on $\lambda$ holds until the $X$ mass grows above 1.3 TeV and kinematically suppress the production rate.

For the dijet channel, we use the CDF~\cite{Aaltonen:2008dn} results of 1.13 fb$^{-1}$ data at 1.96 TeV, which is more constraining for $m_X < 1.2$ TeV in comparison with current LHC data. At the parton level, the only non-trivial cut for our signal events is $|\eta_j|<2.4$. The dijet rate is only dependent on $\lambda_1$. The 95\% C.L. bound is shown in Fig.~\ref{fig:bounds}, with a caveat: The experimental bounds are optimized for particular shape(s) of jet $p_T$ distribution near the resonance. The CDF analysis assumed a heavy vector/fermion as the resonance state. Although there isn't an optimization for our spin-0 $X$, the variation in cross-section due to the $p_T$ shape difference is of the order ${\cal O}(1)$, and we do not expect a qualitative impact on our result. To be conservative, we use the weakest bound from Ref.~\cite{Aaltonen:2008dn}.

For two jets+$\slash{E}_T$, we calculate the combined cross-section from diagrams in both Fig.~\ref{fig:ISGS} and Fig.~\ref{fig:pair_feynman}, then compare with the  two jets + $\slash{E}_T$ exclusive results from ATLAS's multijet + $\slash{E}_T$~\cite{ATLAS-CONF-2013-047} study. We used the same kinematical cut as listed for the `A-Loose' and `A-Medium' signal regions in Ref.~\cite{ATLAS-CONF-2013-047}. For our parton level signal events, $H_T=p_{T}(j_1)+p_{T}(j_2)$ and $M_{\text{eff}}=H_T+\slash{E}_T$. $j_1, j_2$ are the leading and second jet ordered by $p_T$. We found the `A-Medium' region more effective to constrain the $\lambda$s due to less SM background. Both `A-Medium' (solid) and `A-Loose' (dotted) bounds at 95\% C.L. are shown as orange curves in Fig.~\ref{fig:bounds}.

\begin{figure}[h]
\includegraphics[scale=0.6]{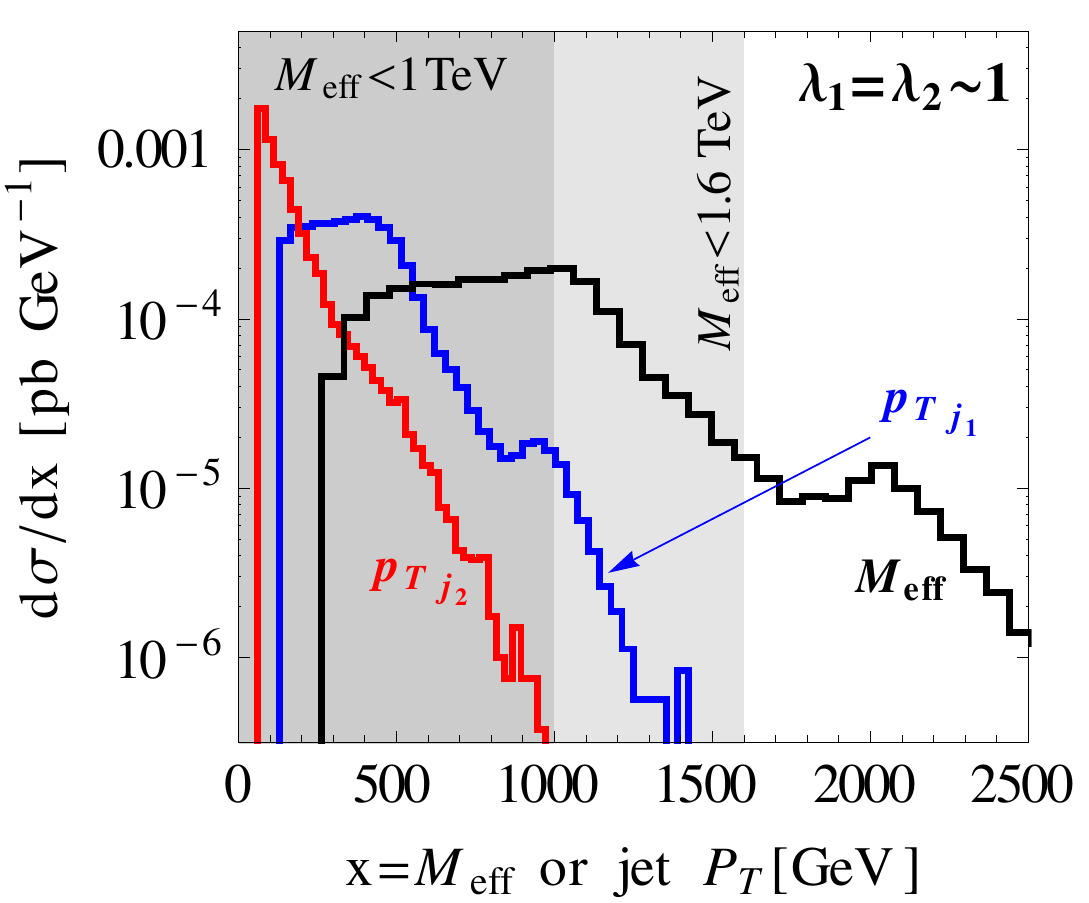} \hspace{0.1cm}
\includegraphics[scale=0.6]{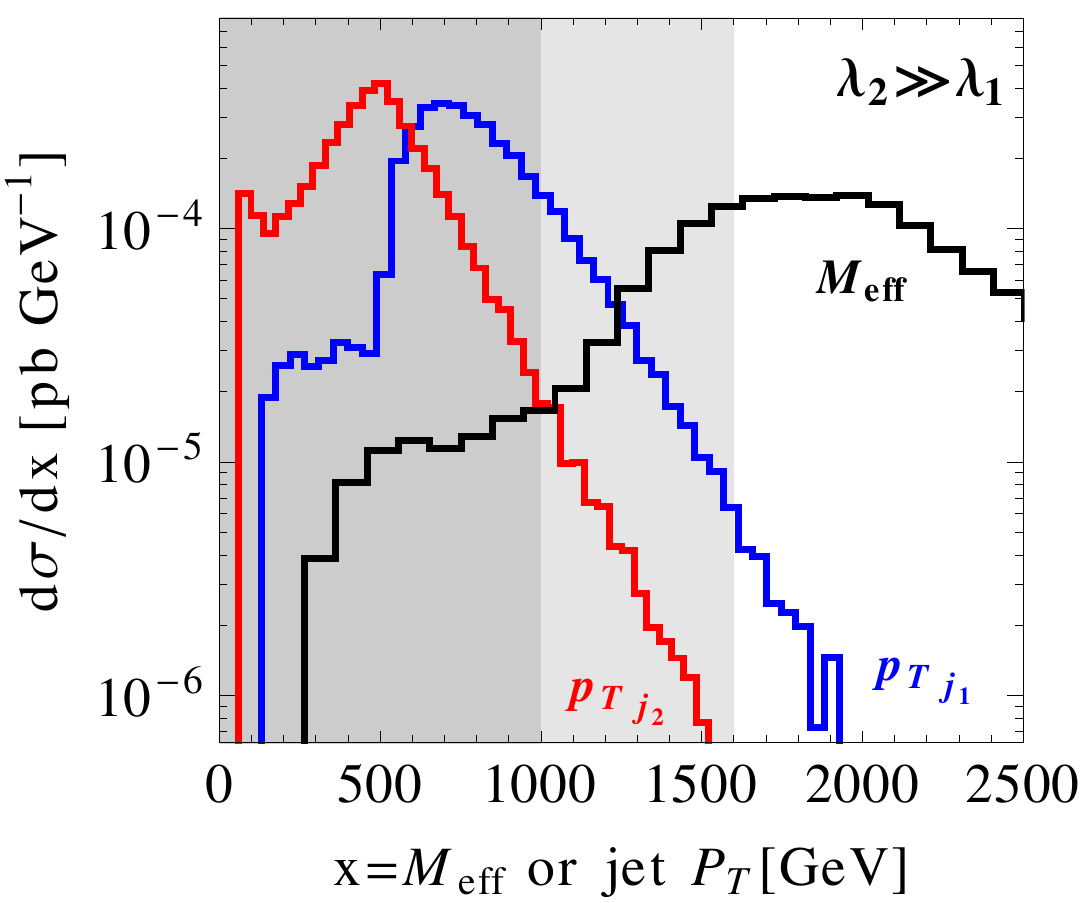}
\caption{Two sample jet $p_T$ (blue and red) and $M_{\text{eff}}$ (black) distributions for $\lambda_1 =\lambda_2 \sim 1$ (left) and $\lambda_2 \gg \lambda_1$ (right). The ISGS process gives a soft secondary jet from gluon splitting, and $M_{\text{eff}}$ near $M_{X1}$. The pair-production process leads to two energy jets and a $M_{\text{eff}}$ peak near $2M_{X1}$. A properly placed $M_{\text{eff}}$ cut can be effective against the ISGS contributions.}
\label{fig:meff}
\end{figure}

Before applying the $M_{\text{eff}}$ cuts, we found for large $\lambda_1$ and $\lambda_2$ the ISGS dominates over pair-production processes in the cross-section. The ISGS cross-section scales as ${|\lambda_1|^2 |\lambda_2|^2}/(2|\lambda_1|^2+|\lambda_2|^2)$ and the shape of its bound resembles that in the monojet case. However, with a large $M_{\text{eff}}$ cut as shown in Fig.~\ref{fig:meff}, or in the case $\lambda_2\gg \lambda_1$, the $u\bar{u}\rightarrow XX^*$ diagram (its $\sigma\propto |\lambda_2|^4$) becomes dominant and bends the combined bound towards a maximally allowed $\lambda_2$. This explains the turn of the 2 jets+$\slash{E}_T$ curves in Fig.~\ref{fig:bounds}.

For paired-dijets, since experimentally two pairs of jets must reconstruct to the same invariant mass, we only consider pair production processes. We use the CMS~\cite{CMS-PAS-EXO-11-016} analysis with 2.2 fb$^{-1}$ data at 7 TeV. Kinematic cuts include $|\eta_j|<2.5$, jet $p_{T}>70$~GeV, the leading jet $p_{T,j1}>140$~GeV and  $\Delta R_{jj}>0.7$. The signal cross-section is calculated as $\sigma=\sigma_{\text{pp}\rightarrow \text{X1X1}}\cdot \text{BR}_{\text{X1}\rightarrow jj}^2\cdot A$, where $A$ is the cut efficiency on the four jet final state that we determine via Monte Carlo. The 95\% C.L. bounds are shown as the black curve in Fig.~\ref{fig:bounds}. 

Interestingly, the paired dijet bound can come with different shapes between the $M_{X1}=500$ GeV and 1 TeV points. The reason, is that the CMS data have a mild ($ 2\sigma\sim 3\sigma$) excess/up-fluctuation at a dijet-invariant mass near 500 GeV, leading to a stronger `observed' local new physics cross-section exclusion bound, which becomes even smaller than the rate of pure QCD production of $XX^*$. In this case, the constraint extends all the way to low $\lambda$ values and a relatively large $\lambda_2/\lambda_1$ rate is needed to suppress the branching ratio into four jets. For the other benchmark $M_{X1}$ at 1 TeV, however, there is no up-fluctuation in CMS data, and the observed bound is weaker than the QCD pair production rate. Only large $\lambda_2\sim {\cal O}(1)$ values become constrained by the NP $dd\rightarrow XX^*$ diagram in Fig.~\ref{fig:pair_feynman}.

\begin{figure}[h]
\includegraphics[scale=0.6]{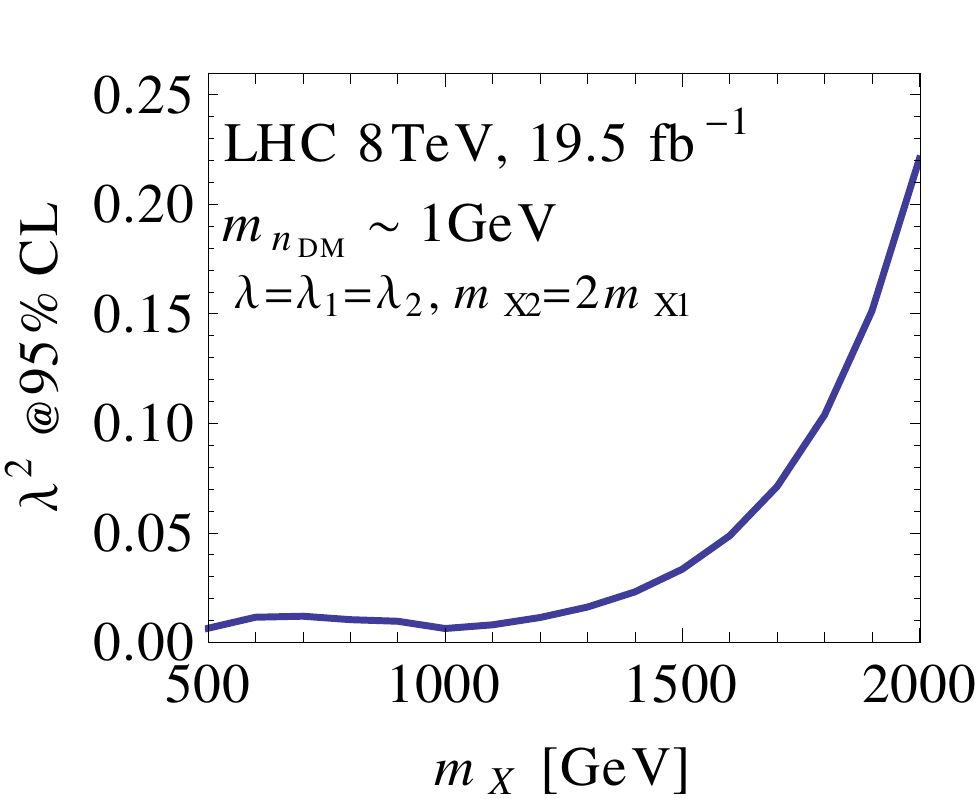}
\caption{Collider results constrain $\lambda (\lambda_1=\lambda_2)$ case) to 0.02 for $m_X$ up to $\sim$TeV. Above TeV, phase space suppression weakens the signal production rate. }
\label{fig:equalscale}
\end{figure}

To summarize, considering only the contribution from one $X$, the monojet channel sets the most stringent constraint on  $\lambda_1$ and $\lambda_2$. With the exception of a narrow window of $M_{X}$ around 500 GeV, one can further set $\lambda \equiv \lambda_1=\lambda_2$ as an over-all coupling scale. Its monojet constraint versus $M_{X1}$ is shown Fig.~\ref{fig:equalscale}. Over a significant mass range below 1.3 TeV, the  
$\lambda$ is constrained to around 0.1.

Although we assume flavor blind couplings in this minimal model, it should be noted the $\lambda$ that couple to the third generation quarks are in principle less constrained due to their subdominant presence from proton PDF, as well as only the light and $b$ quarks being counted as jets in the final state. For instance, the constraints on $\lambda_2$ only apply to first two quark generations. However, the model predicts the $X$ can decay into a dark matter and a top, leading to striking final states of a single or two top(s) + \met process with \met = $M_X/2$. Especially, a mono-top + \met study will provide further constraint on our model.

\section{Discussions}
\label{sect:disc}
We considered a minimal extension of the SM that gives rise to baryogenesis and has a DM candidate of O(GeV) mass.
Two colored scalars with ${\cal O}({\rm TeV})$ mass and a singlet fermion are required in a minimal set up to generate the baryon asymmetry of the universe via renormalizable baryon number violating interactions. The singlet fermion becomes stable and can play the role of a DM candidate, while avoiding rapid proton decay, when it is nearly degenerate in mass with the proton. None of these explanations requires the existence of SUSY.

In this work we investigated the signals and current constraints on this model arising at the LHC. We found the LHC searches, especially with the monojet channel, give significant constraint on the model parameters. For $M_{X1}$ around 1 TeV and below, the lesser between $\lambda_1$ and $\lambda_2$ is constrained to ${\cal O}(0.1)$.  This constraint has impact on the baryon asymmetry calculation, dark matter production and the abundance $Y$ of the decay products of the heavy scalar field whose decay is responsible for generating both baryon and DM abundance. It is interesting to note, if $\lambda_1^1\sim \lambda_1^2$ then the solution to the coincidence problem for $M_{X1}\sim 1$ TeV requires the scalars' mass to be relatively close $M_X \gg \Delta M~(\Delta M\equiv M_{X2}-M_{X1})$, and the contribution from both $X$s should be included. In the non-interference region where $\lambda_1$ is comparable or larger than $\lambda_2$, cross-section doubles and the bounds on $\lambda$ improves by a factor up to 40\%. In Appendix~\ref{sect:scenarios} we list the non-interference scenarios for $\lambda_1\gg\lambda_2, \lambda_1\approx \lambda_2$ where $\Delta M$ can be large, as well as non-interference scenarios in the region $\lambda_1\ll\lambda_2$.  In the latter case, however, interference is more likely to happen:  constructive interference can leads to a more stringent on $\lambda$ bounds up to a factor of 2, which can be considered as an optimistic limit; yet destructive interference may hide signals. With interference the constraints in the $\lambda_1 \ll \lambda_2$ region is highly dependent on the assignment of the complex phase for each $\lambda$.

It should be noted that the most of the cuts in collider searches do not yet optimize for the jet pT distribution in
this model, as indicated in Fig. 4. Improved optimization on monojet pT window can single out the peak structure
at $M_{X1}/2$
, and further enhance the signal to background ratio. 
In the pair-dijet case, a study of $m_{jj}$ from a scalar resonant state can improve the constraint. The correlation between the resonance width and the signal cross section can also be considered as both depend on the new physics coupling.

The measurement of the scalar masses will also lead to the identification of the model at the LHC. However, this constraint can be relaxed if $X d^c_i d^c_j$ is mostly dominated by $Xd^c_ib^c_j$. In this scenario monojet will be less effective to constrain the model, however dijet pair signal will contain  2 b jets which will help us to identify the model at the LHC. 

\bigskip

{\bf Acknowledgements}

This work is supported in part by DOE Grant No. DE-SC0010813.
YG is supported by Mitchell Institute of Fundamental Physics and Astronomy.
TK is also supported in part by Qatar National Research Fund under project NPRP 5-464-1-080. 
\vspace{1cm}

\appendix

\section{$X1$, $X2$ masses and interference}
\label{sect:scenarios}

The baryon abundance that arise from $X$ decay is given in Ref.~\cite{Allahverdi:2013mza},
\bea  
\frac{n_B}{s}&=& \frac{Y_{\cal S}}{8\pi} \frac{1}{M^2_{X2}-M^2_{X1}}
\sum_{i,j,k} \text{Im}(\lambda_1^{1,ij*}\lambda_1^{2,ij}\lambda_2^{1,k*}\lambda_2^{2,k}) \nn \\
&&\times \left[\frac{M_{X1}^2\text{BR}_1}{\sum_{ij}|\lambda_1^{1,ij}|^2+ \sum_{k}|\lambda_2^{1,k}|^2}+\frac{M_{X2}^2\text{BR}_2}{\sum_{ij}|\lambda_1^{2,ij}|^2+ \sum_{k}|\lambda_2^{2,k}|^2}\right],
\eea
where $i,j,k$ are quark generation indices. BR$_1$ and BR$_2$ are the branching ratios of the $X1$ and $X2$ being injected via a late decay of a heavy scalar field ${\cal S}$ that reheats the universe at temperature $T_r$. $Y_{\cal S} = \frac{3 T_r}{4 M_{\cal S}}$ is the dilution factor due to ${\cal S}$ decay. Similarly the dark matter abundance is 
\be
\frac{n_{n_D}}{s}= {Y_{\cal S}}
 \left[\frac{\text{BR}_1 \sum_{k}|\lambda_2^{1,k}|^2}{\sum_{ij}|\lambda_1^{1,ij}|^2+ \sum_{k}|\lambda_2^{1,k}|^2}+\frac{\text{BR}_2 \sum_{k}|\lambda_2^{2,k}|^2}{\sum_{ij}|\lambda_1^{2,ij}|^2+ \sum_{k}|\lambda_2^{2,k}|^2}\right].
 \label{eq:basym}
\ee
 BR$_1$ and BR$_2$ can be free parameters that are completely independent from the baryon number violation process.  As we we see In Setion~\ref{sect:results}, $\lambda_1$ and/or $\lambda_2$ can have a LHC constraint to be $\sim{\cal O}(0.1)$. Assume comparable sizes between imaginary/real parts, we consider the following cases:

 \medskip
 (i)  $|\lambda_1|\gg |\lambda_2|$. Let BR$_1\gg$BR$_2$, $n_B/n_{n_D}=\frac{m_{n_{DM}}}{m_p}\frac{\Omega_B}{\Omega_{n_{DM}}}\sim 0.2$ leads to
 \be 
 \frac{1}{8\pi} \frac{M^2_{X1}}{M^2_{X2}-M^2_{X1}}\frac{\sum_{i,j,k} \text{Im}(\lambda_1^{1,ij*}\lambda_1^{2,ij}\lambda_2^{1,k*}\lambda_2^{2,k})}{\sum_{k}|\lambda_2^{1,k}|^2} \sim 0.2.
 \label{eq:case1}
 \ee
$\Delta M=M_{X2}-M_{X1}$ depends on the ratio between the sum of imaginary parts and $\sum_{k}|\lambda_2^k|$. Since $\lambda_2^2/\lambda_2^1$ is free, this ratio can easily be of the order ${\cal O}(10)$ or larger. When the third fraction in Eq.~\ref{eq:case1} is greater than 3$|\lambda_1|^2$, $\Delta M$ is greater than twice of $\Gamma_X$ and the interference between $X1$ and $X2$ becomes suppressed.

\medskip
 (ii) $|\lambda_1|\sim |\lambda_2|\sim \lambda$. Let BR$_1\sim$BR$_2$, we get
\be 
\frac{n_B}{n_{n_{DM}}}\sim \frac{1}{8\pi}\frac{M_{X1}^2+M_{X2}^2}{M_{X2}^2-M_{X2}^2}\cdot \frac{{C \cdot |\lambda|^2}}{2} = 0.2.
\ee
Similar to (i), denote $C |\lambda|^2$ as the ratio of the sum of the imaginary parts over $\sum_{k}|\lambda_2^k|$. For $C>3$, $\Delta M$ is large than twice of $\Gamma_X$, and the interference can be ignored.

\medskip
 (iii)  $|\lambda_1|\ll |\lambda_2|$. In this case, a small $\lambda_1$ can drive down $\Delta M$, but $\Gamma_X$ still remains large in comparison by $\lambda_2$. We generally expect $\Delta M <\Gamma_X$ and interference occurs. Nonetheless, in the special case that BR$_1\ll$BR$_2$ and $\lambda_2^1 \gg \lambda_2^2$, the large ratio of the $\lambda_2$ terms inside the imaginary parts over  $\sum_{k}|\lambda_2^{2,k}|$ can negate the suppression from the small $\lambda_1$, hence the mass difference can still be large in comparison to $\Gamma_X$ and avoid interference. This requires $|\lambda_2^1/\lambda_2^2| \propto |\lambda_1|^{-2}$.
 
\medskip
As a brief summary, the correct baryon and dark matter densities govern the mass differences between the two heavy colored scalars. When $|\lambda_1|\sim |\lambda_2|$ or $|\lambda_1|\gg |\lambda_2|$ interference can be avoided in collider processes, given an appropriate BR$_1$, BR$_2$ relation and a relatively large sum of the imaginary parts in Eq.~\ref{eq:basym}. However, when $|\lambda_1|\ll |\lambda_2|$, interference occurs except in case both $\lambda_1$ and $\lambda_2^2\ll\lambda_2^1$.

\end{document}